\documentclass{article}

\usepackage{arxiv}

\usepackage[utf8]{inputenc} % allow utf-8 input
\usepackage[T1]{fontenc}    % use 8-bit T1 fonts
\usepackage{hyperref}       % hyperlinks
\usepackage{url}            % simple URL typesetting
\usepackage{booktabs}       % professional-quality tables
\usepackage{amsfonts}       % blackboard math symbols
\usepackage{nicefrac}       % compact symbols for 1/2, etc.
\usepackage{microtype}      % microtypography
\usepackage{lipsum}		% Can be removed after putting your text content
\usepackage{graphicx}
\usepackage[numbers]{natbib}
\usepackage{doi}
\usepackage{multicol}
\usepackage{lineno}
\usepackage{xspace}
\usepackage{changes}
\usepackage{xcolor}

\newcommand{\MR}{MR\xspace}
\newcommand{\MRs}{MRs\xspace}
\newcommand{\selfpulsing}{SP\xspace}

\newcommand{\fig}{Fig.\xspace}

\title{The Taiji microresonator as an unidirectional spiking neuron}

\author{ \href{https://orcid.org/0000-0002-3361-133X}{\includegraphics[scale=0.06]{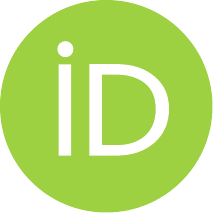}\hspace{1mm}Stefano Biasi}\thanks{Corrisponding author: stefano.biasi@unitn.it.} \\
	Nanoscience Laboratory,\\ 
    Department of Physics,\\
	University of Trento, \\
    38123 Trento, Italy
	\And
	\href{https://orcid.org/0009-0001-3195-8620}{\includegraphics[scale=0.06]{orcid.pdf}\hspace{1mm}Alessandro Foradori} \\
	Photonics Research Group\\
	INTEC-department\\
	Ghent University-imec, \\
    Belgium 
    \And
   \href{https://orcid.org/0000-0002-6112-7457}{\includegraphics[scale=0.06]{orcid.pdf}\hspace{1mm}Riccardo Franchi}\\
	Nanoscience Laboratory,\\ 
    Department of Physics,\\
	University of Trento, \\
    38123 Trento, Italy
    \And
    \href{https://orcid.org/0000-0002-6587-2614}{\includegraphics[scale=0.06]{orcid.pdf}\hspace{1mm}Alessio Lugnan}\\
	Nanoscience Laboratory,\\ 
    Department of Physics,\\
	University of Trento, \\
    38123 Trento, Italy
    \And
	\href{https://orcid.org/0000-0001-6259-464X}{\includegraphics[scale=0.06]{orcid.pdf}\hspace{1mm}Peter Bienstman} \\
	Photonics Research Group\\
	INTEC-department\\
	Ghent University-imec, \\
    Belgium 
    \And
    \href{https://orcid.org/0000-0001-7316-6034}{\includegraphics[scale=0.06]{orcid.pdf}\hspace{1mm}Lorenzo Pavesi}\\
	Nanoscience Laboratory,\\ 
    Department of Physics,\\
	University of Trento, \\
    38123 Trento, Italy
}

\renewcommand{\shorttitle}{\textit{arXiv} Template}

\hypersetup{
pdftitle={A template for the arxiv style},
pdfsubject={q-bio.NC, q-bio.QM},
pdfauthor={David S.~Hippocampus, Elias D.~Striatum},
pdfkeywords={First keyword, Second keyword, More},
}

\begin{document}
\maketitle

\begin{abstract}
While biological neurons ensure unidirectional signalling, scalable integrated photonic neurons, such as silicon microresonators, respond the same way regardless of excitation direction due to the Lorentz reciprocity principle. Here, we show that a non-linear Taiji microresonator is a proper optical analogous of a biological neuron showing both a spiking response as well as a direction dependence response.
\end{abstract}

% keywords can be removed

%\begin{multicols}{2}
Von Neumann computing systems are inherently constrained by the physical separation between processing (CPU) and memory units, resulting in limited speed and energy efficiency.
%bandwidth. Unlike the human brain, these systems struggle to learn and process complex data. 
To address these limitations, neuromorphic systems inspired by biological principles have been proposed \cite{christensen20222022}. These systems use artificial neurons and artificial synapses for processing and memory, with information encoded not only as bits but also as spikes \cite{schuman2022opportunities}. Indeed, spiking neural networks offer reduced energy consumption, intrinsic parallelism, and excellent computational efficiency. Photonics is particularly promising because of its ultra-high speed, low latency, energy efficiency and high bandwidth. Here, integrated optics, especially silicon photonics, could be a disruptive technological platform.  Recent studies have shown that a microresonator (\MR), which leverages material-induced nonlinearity through thermal effects and free carriers, can induce self-pulsing (\selfpulsing) regimes, acting as an energy-efficient artificial spiking neuron \cite{van2012cascadable, xiang2022all, zhang2024chip}.

\begin{figure*}[t!]
\centering
    \includegraphics[scale=1.3]{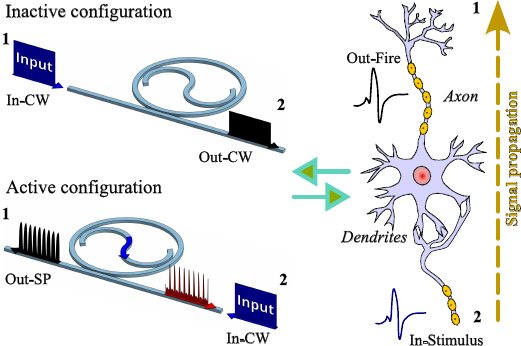}
    \caption{Analogy between a Taiji microresonator (left) and a biological neuron (right). The top shows the inactive configuration with no light coupled in the S-shaped waveguide, while the bottom shows the active configuration with light coupled to the S-shaped waveguide and stored therein. 1 and 2 refer to the left/right ports which can be used as input or output. SP or CW indicate the self pulsed or continuous-wave signal.}
    \label{fig:TaijiNeuron}
\end{figure*}

While a passive \MR can mimic key neuronal functions in the \selfpulsing regimes, it lacks unidirectional signal propagation, showing identical responses when excited from either direction. In contrast, biological neurons 
%are polar, i.e. they 
show unidirectional signal propagation from the dendrites (input ports) to the axon and synaptic terminals (output ports) without the possibility of backtracking (\fig \ref{fig:TaijiNeuron}) \cite{kandel2000principles}. In this work, we demonstrate that a nonlinear Taiji \MR is a proper analog of a biological neuron due to the breaking of the Lorentz  reciprocity.
%While a passive \MR has been shown to mimic key functional properties of biological neurons through \selfpulsing regimes, it fails to achieve the essential unidirectional signal propagation. Specifically, the \MR exhibits the same \selfpulsing responses when excited from left to right or right to left. In contrast, biological neurons naturally exhibit unidirectional signal propagation, with impulses traveling from dendrites to axons without the possibility of backtracking [ref]. In this paper, we experimentally demonstrate how a Taiji \MR can overcome this limitation, making the analogy to biological neurons more precise. 

A Taiji \MR consists of a microring resonator with an embedded S-shaped waveguide (\fig \ref{fig:TaijiNeuron}) \cite{calabrese2020unidirectional, Fan_Taiji22, zeng2023unid}.
The presence of the embedded S-shaped waveguide makes the system non-Hermitian and allows for different behaviors depending on the input port. In the linear regime (low input continuous wave (CW) power), the Taiji \MR behaves like an unidirectional reflector 
\cite{calabrese2020unidirectional}, while in the nonlinear regime (high input power) 
%it can exhibit a different transmission response 
its response strongly depends on the excitation direction, 
thus breaking Lorentz reciprocity \cite{munoz2021nonlinearity}. This is caused by a difference in the energy stored in the \MR because of the presence of the S-shaped waveguide. When the Taiji \MR is excited from port 1 (\fig \ref{fig:TaijiNeuron}, top left), the light coupled to the S-shaped waveguide does not recirculate back to the microring and
%no light is coupled to the S-shaped waveguide and 
it behaves as a usual \MR. On the other hand, when it is excited from port 2 (\fig \ref{fig:TaijiNeuron}, bottom left), the input light in the S-shaped waveguide couples back to the microring. This excites both the propagating and counterpropagating \MR modes with an enhancement of the stored power \cite{munoz2021nonlinearity}. Under CW excitation from port 1, the energy stored in the \MR does not trigger the \selfpulsing regime, and the transmitted light is a constant output signal. We call this working condition the ``inactive configuration’’ of the \MR. Conversely, when port 2 is used as the input port, the energy stored in the \MR triggers the \selfpulsing regime and an oscillating output signal is transmitted by the Taiji \MR. We term this condition, the ``active configuration’’. Therefore, the Taiji \MR shows an unidirectional response in analogy to a biological neuron where the signalling is only possible in one direction. In addition, the Taiji \MR in the active configuration has also other peculiar features such as \selfpulsing states even in reflection (red line in \fig \ref{fig:TaijiNeuron}).

\begin{figure*}[t!]
    \centering
    \includegraphics[scale=1.1]{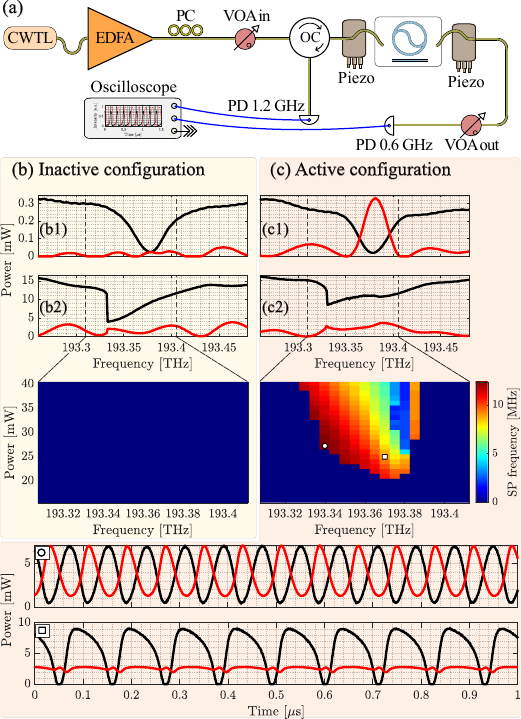}
    \caption{Optical setup (a) and experimental measurements for inactive (b) and active (c) configurations. Spectral responses in linear ((b1), (c1)) and nonlinear ((b2), (c2)) regimes. The color map shows the self-pulsing (SP) frequency as a function of the laser frequency and coupled laser power. The color code is on the rigth. Bottom graphs show the temporal traces of the \MR transmission (black lines) and reflection (red line) when the input has the values given by the circle and square points in the SP map.}
    \label{fig:DesignExp}
\end{figure*}

We demonstrate all these features by using the silicon based Taiji \MR of \cite{biasi2022interferometric} and the experimental setup shown in \fig \ref{fig:DesignExp} (a).
A fiber-coupled continuous wave tunable laser (CWTL) directs light to an Erbium Doped Fiber Amplifier (EDFA), whose output is then passed through a Polarization Control stage (PC) and a Variable Optical Attenuator (VOAin) to set the input power. The light is then routed by an Optical Circulator (OC) into a stripped fiber, which couples it into the sample. At the opposite end, another stripped fiber collects the transmitted light, directing it through a second Variable Optical Attenuator (VOAout) to a photodiode (PD 0.6 GHz). The light reflected by the sample is routed by the optical circulator to another photodiode (PD 1.2 GHz). The signals from both photodiodes are simultaneously acquired with an oscilloscope. The two active and inactive configurations are studied by simply rotating the device within the setup and verifying the same coupling loss in the linear regime.

\fig \ref{fig:DesignExp} (b) and (c) show exemplary results. In the linear regime, ((b1), (c1)), the spectral transmission (black lines) is similar in both configurations, ie. it fulfills the Lorentz reciprocity theorem. However, the presence of the S-shaped waveguide causes the reflection (red lines) to drop to nearly zero in the inactive configuration, while it is very large in the active one \cite{calabrese2020unidirectional}. In the nonlinear regime ((b2), (c2)), both the transmission and reflection spectra differ and do not fulfill the Lorentz reciprocity theorem \cite{munoz2021nonlinearity}. The spectra show the typical nonlinear response with a characteristics triangular shape. The one of the active configuration looks noisier due to the presence of the \selfpulsing. The color maps in \fig \ref{fig:DesignExp} show the \selfpulsing oscillation frequency map, for the input frequency range marked by vertical dashed lines in the spectra. In the inactive configuration, even at a maximum power of 40 mW coupled in the sample, no \selfpulsing is observed. Conversely, in the active configuration, \selfpulsing is observed when the input frequency is resonant with the hot \MR. Examples of the SP signal temporal dependence in transmission (black line) and reflection (red line) are given for a CW input frequency and power corresponding to the points marked by a circle and a square in the \selfpulsing map. Interestingly, two different situations are demonstrated: for the circle point a maximum oscillation frequency (12.4 MHz) and a comparable power between transmission and reflection; for the square point a slow \selfpulsing regime (8.9 MHz) and \selfpulsing mainly in the transmission while the reflection is almost constant. 

These findings shows that the unidirectional properties of nonlinear Taiji \MRs can be used to create photonic spiking neural networks that closely mimic biological systems. A wide range of \selfpulsing regimes modulated by the input parameters is at reach where unidirectionality, inhibition, excitation, refractoriness, and others can be exploited. For example, the unidirectional nonlinear Taiji \MR can be used as a relay neuron in an artificial model of a typical reflex action \cite{kandel2000principles}. In \textit{reflex arc}, a sensory neuron transmits information to a relay neuron, which then relays the signal to a motor neuron. It is crucial that retrograde signaling from the motor neuron does not activate the relay neuron. This is the case with the nonlinear Taiji \MR.

\section{Funding} 
European Union’s Horizon 2020 research and innovation program (grant agreement No. 788793, BACKUP).

\section{Acknowledgments} 
We acknowledge fruitful discussion with Dr. Beatrice Vignoli, Dr. Giovanni Donati and Dr. Antonio Hurtado. S. Biasi acknowledges the cofinancing of the European Union FSE-REACT-EU, PON Research and Innovation 2014–2020 DM1062/2021. A. Lugnan acknowledges funding by the European Union under GA n°101064322-ARIADNE. A. Foradori acknowledges funding by the European Union under GA n°101070238-NEUROPULS.

\section{Data Availability Statement} 
Data available upon valid request.

%\end{multicols}

\bibliographystyle{unsrt}
\bibliography{sample}  %%% Uncomment this line and comment out the ``thebibliography'' section below to use the external .bib file (using bibtex) .





% \end{thebibliography}

\end{document}